# Big Bang, Low Bar – Risk Assessment in the Public Arena

Huw Price[1]

**Abstract:** One of the basic principles of risk management is that we should always keep an eye on ways that things could go badly wrong, even if they seem unlikely. The more disastrous a potential failure, the more improbable it needs to be, before we can safely ignore it. This principle may seem obvious, but it is easily overlooked in public discourse about risk, even by well-qualified commentators who should certainly know better. The present piece is prompted by neglect of the principle in recent discussions about the potential existential risks of artificial intelligence. The failing is not peculiar to this case, but recent debates in this area provide some particularly stark examples of how easily the principle can be overlooked.

## 1. Introduction

From aviation to zoo-keeping, there's a simple rule for safety in potentially hazardous pursuits. Always keep an eye on ways that things could go badly wrong, even if they seem unlikely. The more disastrous a potential failure, the more improbable it needs to be, before we can safely ignore it.

This principle may seem obvious, but it is easily overlooked in public discourse about risk – even, as we'll see, by well-qualified commentators, who should certainly know better. The present piece is prompted by neglect of the principle in recent discussions about the potential risks of artificial intelligence (AI). I don't think the failing is peculiar to this case, but recent debates in this area provide particularly stark examples of how easily the principle can be overlooked.

Part of the problem, in my view, is that there isn't a catchy formulation of this safety principle, already on the tip of educated tongues. By contrast, consider the slogan 'Correlation is not causation.' All scientists, science journalists, and policymakers know this phrase. They thus have a ready reminder of the mistakes that can flow from ignoring the fact in question, and can be held accountable accordingly.

We need a slogan to do the same work in the present case, packaging the principle that the bigger the potential damage, the lower the bar needs to be for taking a risk seriously. My title offers my suggestion – 'Big bang, low bar' – but the field is wide open for better ideas.

## 2. AI and existential risk

First to the AI case. Does AI pose existential risks to humanity? This question has been on some surprising tables lately – tables in the White House and 10 Downing Street, amongst other places, as well as the President's lectern in the European Parliament [1]. These discussions preceded the UK Government's AI Safety Summit, held at Bletchley Park in November 2023 [2].

Some critics feel the issue is getting too much attention. They want to push it aside, or into the distant future, in favour of conversations about the immediate risks of AI. These critics include the editors of *Nature.* A recent *Nature* editorial [3] urges us to '[s]top talking about tomorrow's AI doomsday when AI poses risks today.' The piece concludes: 'Fearmongering narratives about existential risks are not constructive. Serious discussion about actual risks, and action to contain them, are.'

---

[1] Huw Price is Emeritus Bertrand Russell Professor of Philosophy and Emeritus Fellow of Trinity College, Cambridge. He was previously Academic Director of the Centre for the Study of Existential Risk and the Leverhulme Centre for the Future of Intelligence, Cambridge, and a Distinguished Emeritus Professor at the Centre for Science and Thought, University of Bonn.

A week later, in a similar vein, the *Washington Post* published a piece titled '[How elite schools like Stanford became fixated on the AI apocalypse](#)' [4]. This, too, is dismissive about existential risk from AI. It calls it 'a purely hypothetical risk', 'derived from thought experiments at the fringes of tech culture.' It quotes critics who say that these claims 'sound closer to a religion than research', and that their proponents obsess 'over one kind of catastrophe to the exclusion of many others.'

I'll focus on the *Nature* editorial here, for two reasons: it allows me to criticise an institution, rather than named individuals; and *Nature* has a bigger pulpit in science policy than even the most distinguished individual scientists. My message here is that such influence comes with responsibility, not well displayed in this case – a failing linked, apparently, to blindness to the simple principle with which we began. *Nature* thus provides a particularly stark example of the problem I want to discuss. (The *Washington Post* may have an even bigger pulpit, in some senses, but it is not an authoritative voice in science in the same way; and the piece mentioned above was not an editorial, in any case.)

As I said, most of these dismissive opinions about catastrophic long-term risks of AI come from commentators who think that there are plenty of short-term risks, and that we should be worrying about those instead. This means that there's an obvious response, which I'll mention here briefly, and then return to at the end of the piece. Why shouldn't we care about both kinds of AI risks, short- and long-term, and indeed their connections? One connection, plainly, is that responsible governance and regulation of AI needs to deal with both. In the light of this, it would be tying one hand behind our back to focus on one to the exclusion of the other, in either direction. And as Stephen Cave and Seán ÓhÉigeartaigh point out, '[t]echnology can exhibit strong path dependence: decisions people face in the future might be heavily constrained by decisions made now.' [5]

Some groups studying the impacts of AI have been explicit in seeking to be inclusive on this short-term/long-term axis.[2] There are issues of resource allocation, of course. With finite time and money available, choices need to be made. But in my view, this, too, recommends a joined-up approach, to reduce the risk of inefficient specialisation. (Similar points have been made recently by Sætra and Danaher [6].)

The issue of resource allocation brings us back to the 'big bang, low bar' principle. This certainly has implications for how we should allocate time and resources, in managing risks. My main criticism of the *Nature* editorial, and other similarly dismissive responses to the claim that AI involves existential risks, is that they seem blind to these implications. To explain what I mean, let's step back a few paces, and think about risk management in general.

**3. Dodging icebergs**

As I put it above, the safety rule is that we should always keep an eye on ways that things could go badly wrong, even if they seem unlikely. History is full of examples of the costs of getting this wrong: icebergs and frozen O-rings, to mention two famous cases. (I have in mind the *Titanic*, of course, and the Space Shuttle *Challenger*, whose loss in 1986 was caused by effects of cold weather on components called O-rings – more on the latter case below.)

Where does the responsibility lie for enforcing this rule? Sometimes with a single individual, or small group: a ship's captain, say, who decides not to slow down despite warnings about icebergs; or a single NASA committee, deciding to ignore engineers' warnings about the O-rings. But even in these cases, individuals and committees do not operate in a vacuum. It is an institutional responsibility, to some extent, and good institutions are designed with this in mind.

---

[2] This is certainly true of the Cambridge-based Leverhulme Centre for the Future of Intelligence (CFI), of which I was Academic Director from 2016–2021.

For familiar risks, this can be fairly easy. Risk management procedures can themselves be formalised and institutionalised. This is why we have checklists in aircraft cockpits and operating theatres, and triage procedures in an emergency room.

Well-designed safety procedures are sensitive to the costs of errors. This formulation, stressing the issue of responsibility, is from the philosopher Heather Douglas [7].

> In general, if there is widely recognized uncertainty and thus a significant chance of error, we hold people responsible for considering the consequences of error as part of their decision-making process. Although the error rates may be the same in two contexts, if the consequences of error are serious in one case and trivial in the other, we expect decisions to be different. Thus the emergency room avoids as much as possible any false negatives with respect to potential heart attack victims, accepting a very high rate of false positives in the process. … In contrast, the justice system attempts to avoid false positives, accepting some rate of false negatives in the process. Even in less institutional settings, we expect people to consider the consequences of error, hence the existence of reckless endangerment or reckless driving charges.

Douglas goes on to discuss the possibility that '[w]e might decide to isolate scientists from having to think about the consequences of their errors', but rejects it. She argues that 'we want to hold scientists to the same standards as everyone else', and therefore 'that scientists should think about the potential consequences of error.'

For future reference, note something that is especially clear in the case of the justice system: asymmetric consequences of error properly lead to asymmetric burdens of proof. In criminal justice this burden falls more heavily on the side of the prosecution, because of the high costs of false positives.

**4. Dealing with novel risks**

For novel risks these lessons are harder to apply. We don't have the luxury of learning by trial and error, and planning checklists accordingly. But there are some useful guidelines, or rules of thumb, which are based on experience.

One good rule is to listen to experts, insofar as you can tell who they are. By training, experience or simple insight, some people are going to be better sources of advice than others. If those people are warning you about a risk, it makes sense to listen.

A second useful principle is that these experts will be telling you things you don't want to hear, in many cases. That's their job, and not a reason for doubting their expertise. If you find yourself disrespecting the messenger because you don't like the message, that's a danger sign. Good institutions will be designed to mitigate this danger. Hospital managers should not have the power to override their consultants on issues of critical care, no matter how annoying they might find them.

A third useful guideline is that if other people have interests at stake, they, too, may have a reason for disrespecting the messengers. Other parts of an organisation may resent the attention and resources given to the safety team, and people elsewhere may be adversely affected by the team's recommendations.

This means that decision makers often need to act in a noisy environment, where some of the noise is attacks on the character and motivations of the would-be whistleblowers. Real life is messy, but well-designed procedures will try to filter this out, unless it bears on the credibility of the advice.

Finally, it's worth repeating that all of this matters most where the stakes are highest. 'We expect people to consider the consequences of error', as Douglas put it, and the more disastrous the possible consequences, the higher this expectation should be.

**5. Frozen O-rings**

I want to return to AI, but first an example from recent history. On January 28, 1986 the Space Shuttle *Challenger* exploded after launch, killing its seven crew. The explosion was caused by a failure of the O-rings – elastic seals in the solid-fuel booster rockets. This was itself caused by low temperatures the night preceding the launch, a danger that had been foreseen by engineers at the company Thiokol, which supplied these components on the boosters.

One of the engineers was Roger Boisjoly. After his death in 2012, National Public Radio (NPR) [8] recalled an interview they had with him and one of his colleagues, a few weeks after the *Challenger* disaster. Boisjoly told NPR then that they had foreseen the problem with the seals six months previously, and predicted 'a catastrophe of the highest order' involving 'loss of human life' in a memo to managers at Thiokol.

The day before the launch, when the predicted launch-time temperature was below freezing, 'Boisjoly and his four colleagues … concluded it would be too dangerous to launch.' What happened next is a lesson in how not to listen to one's safety team.

> Armed with the data that described that possibility, Boisjoly and his colleagues argued persistently and vigorously for hours. At first, Thiokol managers agreed with them and formally recommended a launch delay. But NASA officials on a conference call challenged that recommendation.
>
> 'I am appalled,' said NASA's George Hardy, according to Boisjoly and our other source in the room. 'I am appalled by your recommendation.' Another shuttle program manager, Lawrence Mulloy, didn't hide his disdain. 'My God, Thiokol,' he said. 'When do you want me to launch – next April?' …
>
> Boisjoly and his colleague … told us that the NASA pressure caused 'Thiokol managers to 'put their management hats on.' …  They overruled Boisjoly and the other engineers and told NASA to go ahead and launch.

**6. Is AGI clear for launch?**

There are differences between the *Challenger* and AI, of course. One is that for the *Challenger*, the issue of expertise was exceptionally clearcut. Boisjoly and his fellow engineers had actually designed the system whose failure they predicted.

In the AI case the community now concerned about existential risks includes many leading scientists in the field, both from academia and industry.[3] These concerns mostly focus on the risks of so-called artificial general intelligence (AGI) – machines that reach or exceed human cognitive abilities on a wide range of tasks.[4] But AGI is not launching tomorrow. So the link between the engineers and the (claimed) risky technology is not nearly so direct.

---

[3] In March 2023, many of them signed an open letter [9], calling for a six-month pause on the training of AI systems more powerful than GPT-4.

Another difference is that in the *Challenger* case there were institutions in place, with the authority to assess the advice of the engineers. Those institutions failed, apparently for reasons that seem familiar, in the light of our simple rules. Appalled disdain for one's experts is not a helpful frame of mind. Thiokol's management hat should not have overruled its engineering hat. Nevertheless, the institutions were there. We know whose responsibility it was to get things right.

In the AI case, by contrast, we don't yet have institutions to do the job. Much of the recent discussion has been about how to create them. In our view, this means that the responsibility that such institutions would bear has to fall much more broadly, for the time being. It has to rest on the shoulders of other institutions and individuals, scientific and otherwise, who are capable of influencing discussion in this case. As the world's leading scientific journal, *Nature* is one of these institutions, in my view.

Now to the biggest difference. In the case of the *Challenger,* the predictable cost of a false negative was already very high. The Space Shuttle had no crew recovery system, so a loss of the vehicle meant loss of the crew. Still, these losses are tiny compared to those that would be at stake if AI really comes with existential risks. Seven people compared to eight billion, not to mention the future of the biosphere.

What level of safety should we be aiming for? In the case of the *Challenger*, NASA management told the Rogers Commission that they aimed for a risk of catastrophic failure of 1 in 100,000. In Richard Feynman's appendix to the commission's report [12], he noted that NASA engineers he surveyed estimated the risk much higher – about 1 out of 100 launches.

Obviously, we should be looking for a much safer comparison for the future of AI. Perhaps commercial aviation? There are approximately 100,000 commercial flights per day, apparently. If we assume about 3 serious accidents per year, that's about 1 serious accident per 10 million flights. We might regard that as a very crude upper bound on the level of risk that would be acceptable for species-threatening AI. The potential bang is very, very big, and so the bar for taking the risk seriously is extremely low.

**7. Two objections**

It may seem that another difference between the *Challenger* and AI is that in the former case, the engineers could rely on 'objective' evidentially-grounded probabilities. Where could such probabilities come from, in the AI case? But this difference is one of degree. There were no large statistical trials to draw on in the *Challenger* case, obviously. More importantly, it is common and unavoidable in risk-management, and cost-benefit analysis generally, that the probabilities needed are often imprecise, contested, and necessarily extracted from qualitative assessments by experts [13]. There is no alternative to working with these rough materials, whenever a choice cannot be avoided – whenever 'doing nothing' is simply one option amongst others.

Another objection, perhaps to be detected in NASA's disdainful response to Boisjoly and his colleagues, is what's known in the existential risk literature as 'Pascal's mugging' [14]. If tiny probabilities matter, won't the 'big bang, low bar' principle prohibit almost anything, if the claimed bang is sufficiently big? ('When do you want us to launch – never?') This is a reasonable point, and any practical response will need to lean on the expertise of those who are raising the concern, as we have already recommended. But the objection isn't really relevant to the present situation in AI, where claimed probabilities of disaster are very far from tiny [15].

---

[4] The main concern is the so-called control problem, or alignment problem, that an AGI might find unintended ways of achieving its assigned goals, with disastrous consequences for humans [10]. It has been noted that less powerful AI might also pose existential risks, in other ways [11].

**8. Reckless endangerment?**

As I said, these concerns about AI have been raised by some (though by no means all) of the field's leading scientists. These are not fringe figures. One of leading voices [10] is Professor Stuart Russell (Berkeley), who was the BBC's Reith Lecturer in 2021 – that's a long way from the fringes, by anybody's standards.

It seems clear that *Nature's* response does not rest on a comparable body of expertise, assembled to support a case on the other side. This is not to take a side on the issues, but simply to point out that *Nature* cannot possibly have had the expertise required to do so, let alone at the level of certainty that would be required. Yet they dismiss the concerns of Russell and his colleagues as '[f]earmongering narratives' (to be contrasted, in their view, with the kind of '[s]erious discussion about actual risks' for which they express support).

The editorial also impugns the motives of scientists involved, suggesting that their concerns are 'a magician's sleight of hand, [to draw] attention away from the real issue[s]'; and that 'the spectre of AI as an all-powerful machine … works to the advantage of tech firms: it encourages investment and weakens arguments for regulating the industry.'

Dismissiveness of this kind would be surprising in an ordinary scientific dispute, especially given the academic standing of some of the targets. But in a case in which the bar for taking matters seriously is properly so low, it is little short of reckless. The most charitable explanation seems to be that *Nature's* editors have missed the 'big bang, low bar' principle altogether.

I want to stress that I don't take *Nature* to be uniquely at fault. As I said above, I have focussed on their editorial for two reasons: it is easier to criticise an institution than named individuals, and *Nature* is (to my knowledge) the most influential voice to take this dismissive attitude. But as the editorial says, it was endorsing viewpoints that were already in play. A number of academics have made similarly dismissive public remarks about claimed existential risks of AI, and *Nature* notes that it had spoken to some of them.

Some of the scholars in this latter camp certainly do have relevant expertise in the field – there are major figures on both sides. However, there is an important asymmetry in the debate, which is easily overlooked. Because the cost of a false negative (i.e., of wrongly setting aside concerns about catastrophic risks) is so high, the bar is rightly much lower on one side of the debate than on the other. This is the burden of proof point we noted earlier, and it is a straightforward consequence of the 'big bang, low bar' principle. If someone objects that the asymmetry is unfair to one side of the debate, then they haven't understood this basic principle of risk management.

Thus the bar for making an adequate case for *setting aside* concerns about existential risk from AI is exceedingly high. For this reason, I think we should at present be sceptical of those who are overly dismissive of claims about such risks, even if they have expertise in the field themselves. Have they really made a realistic estimate of the possibility that they themselves might be wrong, and set it against a proper assessment of the degree of certainty that a risk of this magnitude requires? If not, then the 'big bang, low bar' principle is again being ignored.

## 9. Not just AI

As I said at the beginning, the AI case is unlikely to be unique. Other technologies are likely to generate new kinds of risks, some of them probably catastrophic. AI itself may be an enabler, in some such cases, as many authors have noted. This means that future public debates are likely to be treading similar ground. It would be helpful if we could raise the general level of discussion, by making it harder for prominent voices to make such important mistakes.

I noted that novel risks often need to be assessed in noisy environments, without clear lines of responsibility. In these circumstances, authoritative scientific voices – national academies, for example, as well as respected international journals – have a useful role to play. They can stand back from the noise, and try to improve the quality of the debate. Among other things, they could remind all sides of the basic principles of risk management, including the one whose absence is so striking in the case we have been discussing.

Meanwhile, my recommendation for the AI case is a simple one, at least at the relevant strategic level. The challenges of mitigating the risks of AI, and achieving its benefits, will be hugely complicated, needing expertise from many directions. These challenges will be best addressed by a community which is inclusive on many dimensions, including the short-term/long-term dimension. Recent polarisation has been a wholly unnecessary distraction from some of the most important issues of our time.

## Acknowledgements

The present piece is a descendant of a preprint co-authored with Matthew Connolly [16], which also resulted in a joint Letter to the Editor in *Nature* [17]. I am very grateful to Professor Connolly for his permission to reuse the material here, and also to Haydn Belfield, Martin Rees, and Julian Huppert, for comments and discussion. I am also grateful for comments from two anonymous referees for *Royal Society Open Science.*